\documentclass[prd,showpacs,showkeys,aps,amsmath,amssymb]{revtex4}
\usepackage{amsmath,amssymb}
\usepackage{graphicx}
\usepackage{dcolumn}
\usepackage{bm}
\usepackage{tabularx}

   \def\Lr{{\hat r}(1)}
  \def\Rr{{\hat r}(2)}

  \def\vecr{\mbox{\boldmath $r$}}
  \def\vectau{\mbox{\boldmath $\tau$}}
  \def\vecR{\mbox{\boldmath $R$}}
    \def\vecL{\mbox{\boldmath $L$}}
    \def\vecT{\mbox{\boldmath $T$}}
        \def\vecI{\mbox{\boldmath $I$}}
  \def\Ld{{\hat\delta{(1)}}}
  \def\Rd{{\hat\delta{(2)}}}
  
  \def\tr{{\rm tr}}
  \def\gsl#1{\rlap{\slash}#1} 
   
  \def\p{\gsl p} 
   
  \def\epsdel(#1,#2,#3,#4){\quad(\delta_{#1#3}\delta_{#2#4}
                                 -\delta_{#1#4}\delta_{#2#3})} 
  \def\<>#1{\langle#1\rangle} 
  \def\><{\rangle\langle} 
\def\ben{\begin{equation}}
\def\een{\end{equation}}
\def\bey{\begin{eqnarray}}
\def\eey{\end{eqnarray}}
\def\ba{\begin{array}}
\def\ea{\end{array}}
\def\benmrt{\begin{enumerate}}
\def\eenmrt{\end{enumerate}}

\def\psla{p{\raise1pt\hbox{$\!\!/$}}}
\def\dsla{\partial{\raise1pt\hbox{$\!\!\!/$}}}
\def\Dsla{D{\raise1pt\hbox{$\!\!\!/$}}}
\def\xsla{x{\raise1pt\hbox{$\!\!\!/$}}}

\def\vecr{\mbox{\boldmath $r$}}

\def\jmu5{j_{\mu 5}^{(i)}(0)}
\def\jnu5{j_{\nu 5}^{(i)}(0)}

\def\qq0v{\langle0\!\mid\!{\bar q}q\!\mid\! 0\rangle}

\def\qc0f{\langle0\!\mid\!{\bar q}q\!\mid\!0\rangle_{F}}

\def\qsq0f{\langle0\!\mid\!{\bar q}\sigma_{\mu\nu}q\!\mid\!0\rangle_{F}}

\def\qgdq0f{\langle0\!\mid\!{\bar q}{\cal 
S}\gamma_{\mu}D_{\nu}q\!\mid\!0\rangle_{F}}

\def\qddq0f{\langle0\!\mid\!{\bar q}{\cal
S}D_{\mu}D_{\nu}q\!\mid\!0\rangle_{F}}

\def\3mmtm{|{\bf q}|^2}

\def\eq#1{Eq.(\ref{#1})}
\def\eqs#1#2{Eqs.(\ref{#1}) and (\ref{#2})}

\def\Ref#1{[\ref{#1}]}
\def\Refs#1#2{[\ref{#1},\ref{#2}]}

\def\p0{p_0}

\def\gam3{\mbox{\boldmath{$\gamma$}}}
\def\e0{E_{0}(s_{0},s)}
\def\e1{E_{1}(s_{0},s)}
\def\e2{E_{2}(s_{0},s)}

\def\aplt{\kern0.3333em \raise 0.2ex \hbox{$<$}%
\kern-0.8em \lower0.8ex \hbox{$\sim$}%
\kern0.3333em}
\def\aplg{\kern0.3333em \raise 0.2ex \hbox{$>$}%
\kern-0.8em \lower0.8ex \hbox{$\sim$}%
\kern0.3333em}

\begin{document}

\preprint{}
\title{$ppK^-$ bound states from Skyrmions}

\author{Tetsuo NISHIKAWA}
\email{nishi@th.phys.titech.ac.jp} 
\affiliation{%
Department of Physics, Tokyo Institute of Technology, 2-12-1, Oh-Okayama, Meguro, Tokyo 152-8551, Japan
}
\author{Yoshihiko KONDO}
\email{kondo@kokugakuin.ac.jp} 
\affiliation{%
Kokugakuin University, Higashi, Shibuya, Tokyo 150-8440, Japan
}

\date{\today}

\begin{abstract}
The bound kaon approach to the strangeness in the Skyrme model is applied to investigating the possibility of deeply bound $ppK^-$ states.
We describe the $ppK^-$ system as two-Skyrmion around which a kaon field fluctuates. 
Each Skyrmion is rotated in the space of SU(2) collective coordinate.
The rotational motions are quantized to be projected onto the spin-singlet proton-proton state.
We derive the equation of motion for the kaon in the background field of two Skyrmions at fixed positions.
From the numerical solution of the equation of motion, it is found that the energy of $K^-$
can be considerably small, and that the distribution of $K^-$ shows molecular nature of the $ppK^-$ system.
For this deep binding, the Wess-Zumino-Witten term plays an important role.
The total energy of the $ppK^-$ system is estimated in the Born-Oppenheimer approximation. The binding energy of the $ppK^-$ state is $B.E.\simeq 126$ MeV.
The mean square radius of the $pp$ subsystem is $\sqrt{\langle r_{pp}^2\rangle}\simeq 1.6$ fm. 
\end{abstract}
\maketitle

\section{introduction}
For recent years, lots of theoretical or experimental efforts to explore the possibility of nuclear $\bar K$-bound states \cite{AY} have been made.
Although no firm evidences to show their existence are known up to now,
there is one result reported by FINUDA collaboration \cite{FINUDA} which {\it may} suggest the existence of the lightest nuclear $\bar K$-bound state, $ppK^-$.
In the experiment at DA$\Phi$NE, it was observed 
that $\Lambda$ and $p$ from $K^-$ absorption on ${}^6$Li, ${}^7$Li and ${}^{12}$C at rest has a strong back-to-back correlation,
and that the invariant mass spectrum of $\Lambda$ and $p$ shows a peak.
The collaboration advocates that the observation can be interpreted as a signal of the formation of deeply bound $ppK^-$ state, whose binding energy is $115\,{\rm MeV}$ and the width is $67\,{\rm MeV}$. 

This experiment is motivated by the idea proposed by Akaishi and Yamazaki (AY) \cite{AY} suggesting the existence of deeply bound $\bar K$ nuclei.
It is based on the assumption that $\Lambda(1405)$ baryon is a ${\bar K}N$ bound state formed by the strong attraction in $I=0$ ${\bar K}N$ channel.
One may then expect that when a $K^-$ is injected in a nucleus it may attract surrounding $p$'s to form a shrunk nucleus.
The $K^-$ is bound deeply so that an absorption reaction, $K^-p\rightarrow \pi\Sigma$, is energetically closed, and accordingly, it can have a long life time in a nucleus.

However, it has not yet been established that the peak observed by FINUDA really corresponds to the state proposed by AY. 
Magas {\it et al.} \cite{magas} claimed that the peak corresponds mostly to the process $K^-pp\rightarrow\Lambda p$ followed by final-state interactions of the produced particles with the daughter nucleus.
Even if we suppose the peak to be a $ppK^-$ bound state, it is strange that it is much deeper than the original AY prediction \cite{AY}: the binding energy $B.E.=48\,{\rm MeV}$ and the width $\Gamma=61\,{\rm MeV}$.
Recently, two groups \cite{gal,ikeda} performed Faddeev calculations.
The authors in Ref. \cite {gal} obtained $B.E.=55-70\,{\rm MeV}$ and $\Gamma=95-110\,{\rm MeV}$. The result in Ref. \cite{ikeda} are $B.E.\sim 80$MeV and $\Gamma\sim 73$MeV. They are at odds with both of the result by FINUDA and AY prediction. On the other hand, an attempt to describe the $\bar K$-nuclei as $\Lambda(1405)$-heypernuclei has also been made in Ref.\cite{yasui}.

A new experiment are planned to be performed at J-PARC searching for the deeply bound $ppK^-$ state by the missing-mass spectrum of the ${}^3$He(in-flight $K^-$, $n$) reaction, together with the invariant-mass spectra detecting all particles decaying from the $ppK^-$ bound state. 
It would be naively expected that a clear signal for formation of kaonic nuclei appears in this measurement, since a lighter nucleus is chosen as a target. 
Indeed, it has been suggested theoretically that a distinct peak of the $ppK^-$ bound state 
can be observed in the spectrum of the ${}^3$He(in-flight $K^-$, $n$) reaction
if some conditions for the $ppK^-$ optical potential are satisfied \cite{koike}.
 
Our interests in this paper are whether the deeply bound $ppK^-$ state can be realized in the context of the topological soliton model of baryons, the Skyrme model \cite{skyrme}. 
For this purpose, we employ the bound kaon approach to the strangeness in the Skyrme model \cite {CK}, which describes hyperons as the bound states of an anti-kaon and a topological soliton of the pion field (\lq\lq Skyrmion").

The Skyrme model is a low energy effective theory of QCD at large number of colors.
In the limit of large number of colors, as was shown by t'Hooft \cite{thooft}, QCD reduces to a theory of weakly interacting meson theory.
The action of the Skyrme model is a chiral effective theory written in terms of the Nambu-Goldstone boson fields.
Nucleons emerge as topological solitons of the ${\rm SU(2)_f}$  sector in the Skyrme Lagrangian \cite{witten1,adkins}.

One of the way to introduce the strangeness to the model is to assume a kaon field fluctuating around the SU(2) Skyrmion (bound kaon approach).
One finds the existence of bound states of the kaon and a Skyrmion, which can be identified to be hyperons. 
The lowest bound state has the quantum number $l=1$, $t=1/2$, 
where $l$ is the orbital angular momentum of the kaon and $t$ the combined angular momentum and isospin, $\vecT=\vecL+\vecI$,
respectively. 
The parity of the $l=1$, $t=1/2$ state is totally positive, which is assigned to positive parity hyperons. 
A notable feature is the presence of a bound state in the negative parity state, $l=0$, $t=1/2$, which lies above the $l=1$, $t=1/2$ state. This state probably corresponds to $\Lambda(1405)$ baryon.
Whereas the constituent quark models have difficulties to describe $\Lambda(1405)$, this approach predicts the static properties of $\Lambda(1405)$ \cite{sco} as well as octet and decouplet baryons in good agreement with the empirical values.
In addition, an interaction originating from the WZW term acts on the $S=+1$ state, {\it e.g.} pentaquark, repulsively, and the state is pushed away into the continuum, while the interaction acts attractively on the $S=-1$ state, leading to the formation of the bound states.


Thus, the bound kaon approach is a theory naturally describing both of the positive-parity hyperons and the lowest negative parity state, $\Lambda(1405)$, on the same ground.
It is also worth mentioning that this approach has no parameter once we adjust $F_\pi$ and $e$ (for their definitions, see below) to fit the $N$ and the $\Delta$ masses in $\rm SU(2)_f$ sector. 
Therefore, the ${\bar K}N$ interaction, which is a key ingredient for the study of $\bar K$ nuclei, is unambiguously determined.
In these respects, it is of great significance to investigate the issue of the exotic nuclei such as $ppK^-$, $K^-ppn$, $K^-pnn$ and so on in the context of the bound kaon 
approach to the Skyrme model.

We describe the $ppK^-$ system as two-Skyrmion around which a kaon field fluctuates. Each Skyrmion is rotated in the space of collective coordinate and 
its rotational motion is quantized to be projected onto a relevant two-nucleon state.
We adiabatically treat the nucleon-nucleon radial motion
and derive the kaon's equation of motion when the position of the Skyrmions are fixed first.
Then we obtain the energy of kaon as a function of the relative distance between the two Skyrmions, which tells us whether the kaon can be deeply bound to two-proton or not.
For the existence of nuclear $\bar K$ bound states, it is necessary that $\bar K$
gains sufficiently large binding energy in nuclei.
If such nuclei exist, the nuclear components rearrange themselves under the influence of the strong attraction in $I=0$ $\bar K$$N$.
Then the nuclear part in $\bar K$ nuclei is excited relative to the original nuclear system. Therefore, the energy gained by $\bar K$ must be large enough to compensate the energy loss of nuclear component and to deeply bind the total system.  
If a strong binding of $K^-$ is possible, it is also an interesting subject
how the mechanism responsible for the strong binding is explained in the solitonic picture of baryons. 

In our previous paper \cite{NK}, we have presented the derivation of the kaon's equation of motion and its numerical solution.
It was shown that a strong binding of $K^-$ to $pp$ can occur.
Needless to say, it cannot be taken as an evidence for the actual deep binding of the $ppK^-$ system until the two-proton dynamics is treated.
In the present paper, we give a detailed description of our approach, and attempt to  solve the dynamics of the radial motion of the protons under the strong attaractive force mediated by $K^-$, in addition to the ordinary nuclear force.
Then we can estimate the binding energy of the total $ppK^-$ system within the Born-Oppenheimer approximation.
The possible structure of the $ppK^-$ state is also discussed.

The organization of this paper is as follows. In the second section, we derive the kaon's equation of motion, and show its numerical solutions in the third section.
We solve the radial motion of the two-proton in the forth section.
Discussion and summary are given in the fifth section. Full expression of the kaon Lagrangian and the usefull formulae for collective coordinate quantization are gathered in the appendices.

\section{derivation of the kaon's equation of motion}
\label{deriveom}
Let us begin with showing how the $K^-$ coupled to $pp$ is described in the bound kaon approach to the Skyrme model.
We consider two Skyrmions fixed at positions with the relative distance, $R$,
and assume the presence of the kaon field fluctuating around the Skyrmions.
The equation of motion for the kaon in the background field of the Skyrmions is then derived, from which we know the behavior of the $K^-$ coupled to $pp$.

The action of the Skyrme model is given by
\bey
\Gamma&=&\int d^4 x\left\{\frac{F_\pi^2}{16}\tr(\partial_\mu U^\dagger\partial^\mu U)
+\frac{1}{32e^2}\tr\left[\partial_\mu UU^\dagger,\partial_\nu UU^\dagger\right]^2
\right\}
+\Gamma_{\rm SB}
+\Gamma_{\rm WZW}
\label{action}
\eey
where $U$ is the chiral SU(3) field built out of the eight Nambu-Goldstone bosons.
$\Gamma_{\rm SB}$ is the symmetry breaking term \cite{sco} given by 
\bey
\Gamma_{\rm SB}&=&\int d^4 x
\left\{\frac{F_\pi^2m_\pi^2+2F_K^2m_K^2}{48}\tr\left[U+U^\dagger-2\right]
+\frac{F_\pi^2m_\pi^2-F_K^2m_K^2}{24}\tr\left[\sqrt{2}\lambda_8\left(U+U^\dagger\right)\right]\right.
\cr&&\left.
-\frac{F_\pi^2-F_K^2}{48}\tr\left[(1-\sqrt{3}\lambda_8)(U\partial_\mu U^\dagger\partial^\mu U+U^\dagger\partial_\mu U\partial^\mu U^\dagger)\right]\right\},
\label{sbterm}
\eey 
where $m_{\pi(K)}$ and $F_{\pi(K)}$ are the mass and the decay constant of the pion (kaon), respectively.
The last term in \eq{sbterm} has a role to renormalize the kinetic energy term for the kaon, while the first two terms renormalize the mass term.
$\Gamma_{\rm WZW}$ is the Wess-Zumino-Witten anomaly action \cite{witten2}:
\begin{eqnarray}
\Gamma_{WZW}=-{iN_c\over240\pi^2}\int d^5x \epsilon^{\mu\nu\alpha\beta\gamma}
\tr \left(U^\dagger \partial_\mu UU^\dagger \partial_\nu UU^\dagger \partial_\alpha UU^\dagger \partial_\beta UU^\dagger \partial_\gamma U\right).
\end{eqnarray}
$N_c$ denotes the number of colors.

We assume the following \lq\lq product" ansatz for the chiral field representing $KNN$ system,
\ben
U=U(1)U_KU(2),
\label{ansatz}
\een
where $U(1)$ and $U(2)$ are the fields of the baryon number $B=1$ SU(2) Skyrmions located at
$\vecr(1)=\vecr-{\vecR}/{2}$ and $\vecr(2)=\vecr+{\vecR}/{2}$, respectively.
Their explicit expressions are as follows,
\ben
U(i)=\left(\begin{array}{cc}
u(i)&\bf{0}\\
\bf{0}&0
\end{array}\right),\,\,
u(i)=e^{iF(r(i))\vectau\cdot{\hat\vecr(i)}},\,\, (i=1,2),
\een
where $r(i)=|\vecr(i)|$, ${\hat\vecr(i)}=\vecr(i)/r(i)$, $F(r)$ is the profile function of an isolated Skyrmion \cite{adkins} and $\vectau$ the Pauli matrices.
$U_K$ is the field carrying strangeness. Its form is
\bey
U_K={\rm exp}\left[i\frac{2\sqrt{2}}{F_K}\left(
\begin{array}{cc}
\bf{0}  & K  \\
K^\dagger  & \bf{0}  
\end{array}
\right)\right],
\eey
where $K$ is the usual kaon isodoublet,
\bey
K=\left(
\begin{array}{c}K^+ \\ K^0\end{array}\right).
\eey
Each Skyrmion is rotated in the space of SU(2) collective coordinate, $A_1$ or $A_2$,
as
\ben
u(1)\rightarrow A_1u(1)A^\dagger_1,\quad u(2)\rightarrow A_2u(2)A^\dagger_2.
\label{rotate}
\een

By substituting the ansatz, \eq{ansatz}, with the replacement \eq{rotate} into the action, \eq{action}, we obtain the Lagrangian for the kaon field in the presence of the background $B=2$ Skyrmion.
After expanding the Lagrangian up to the seceond order in $K$, we obtain
\begin{eqnarray}
{\cal L}&=&
(D_\mu K)^\dagger D^\mu K-m_K^2K^\dagger K
-{1\over8}K^\dagger K\left\{\tr(\partial_\mu U_{BB}^\dagger\partial^\mu U_{BB})
+{1\over e^2F_K^2}\tr[\partial_\mu U_{BB}U_{BB}^\dagger,\partial_\nu U_{BB} U_{BB}^\dagger]^2\right\}
\cr&&
-{1\over e^2F_K^2}\left\{2(D_\mu K)^\dagger D_\nu K\tr(A^\mu A^\nu)
+{1\over2}(D_\mu K)^\dagger D^\mu K\tr(\partial_\nu U_{BB}^\dagger\partial^\nu U_{BB})
-6(D_\mu K)^\dagger[A^\nu,A^\mu]D_\nu K\right\}
\cr&&-{iN_c\over F_K^2}B^\mu[K^\dagger D_\mu K-(D_\mu K)^\dagger  K],
\label{lag0}
\end{eqnarray}
where $U_{BB}$ represents the product of rotating solitons,
\ben
U_{BB}=A_1u(1)A^\dagger_1A_2u(2)A^\dagger_2.
\label{ubb}
\een
In \eq{lag0}, the \lq\lq covariant derivative" $D_\mu$ is defined by
\ben
D_\mu K=\partial_\mu K+V_\mu K
\een 
and 
\ben
V_\mu=[L_\mu(1)+R_\mu(2)]/2,\quad A_\mu=[L_\mu(1)-R_\mu(2)]/2,
\label{VA}
\een
where
\bey
L_\mu(1)=A_1u^\dagger(1)A_1^\dagger\partial_\mu[A_1u(1)A_1^\dagger], 
\quad R_\mu(2)=A_2u(2)A^\dagger_2\partial_\mu[A_2u^\dagger(2)A^\dagger_2].
\eey
The last term in \eq{lag0} comes from the WZW term and 
$B^\mu$ is the baryon number current given by 
\bey
B^\mu={\epsilon^{\mu\nu\alpha\beta}\over24\pi^2}\tr(U_{BB}^\dagger\partial_\nu U_{BB} U_{BB}^\dagger\partial_\alpha U_{BB} U_{BB}^\dagger\partial_\beta U_{BB}).
\label{bmu}
\eey
Here we note that the Lagrangian, \eq{lag0}, has the same form as that for $B=1$ Skyrmion \cite{CK} except that the background Skyrmion is not the single $B=1$ Skyrmion but the product of $B=1$ Skyrmions, \eq{ubb}.
It should be also noted that the $KNN$ interaction is unambiguously determined 
as in \eq{lag0} once the ansatz for $U$ is given.

We neglect the terms suppressed by $1/N_c$ in \eq{lag0}.
Since the time-derivative of the collective coordinate is $\dot{A}_{1,2}\sim{\cal O}(N_c^{-1})$, we see $A_0$, $V_0$ and $B_j$ are ${\cal O}(N_c^{-1})$ from their definitions, \eqs{VA}{bmu},\
\ben
A_0,V_0\sim{\cal O}(N_c^{-1}), \quad B_j\sim{\cal O}(N_c^{-1}).
\een
Then the Lagrangian for the kaon field up to $O(N_c^{0})$ terms reads as follows,
\bey
{\cal L}
&=&(\partial_0K)^\dagger \partial_0K\left[1+{1\over 2e^2F_K^2}\tr(\partial_j U_{BB}^\dagger\partial_j U_{BB})\right]
+K^\dagger D_jD_jK-m_K^2K^\dagger K
\cr&&-{1\over8}K^\dagger K\Bigg\{-\tr(\partial_j U_{BB}^\dagger\partial_j U_{BB})
+{1\over e^2F_K^2}\tr[\partial_j U_{BB}U_{BB}^\dagger,\partial_i U_{BB} U_{BB}^\dagger]^2\Bigg\}
\cr&&+{1\over e^2F_K^2}\Bigg\{
2K^\dagger D_j[D_i K\tr(A_j A_i)]
+{1\over2}K^\dagger D_j[D_jK\tr(\partial_iU_{BB}^\dagger\partial_iU_{BB})]
-6K^\dagger D_j([A_i,A_j]D_iK)\Bigg\}
\cr&&-{iN_c\over F_K^2}[K^\dagger \partial_0K
-(\partial_0K)^\dagger K]B^0 +O(N_c^{-1}).
\label{lag}
\end{eqnarray}

A comment is in order here. 
It should be noted that in the Lagrangian \eq{lag} $U_{BB}$ must be the product not of the static solitons but of the rotating solitons, \eq{ubb}.
In the case of $B=1$, the effect of rotation is suppressed in the limit of large $N_c$ since the collective coordinate enters the Lagrangian only through its time-derivative, which is ${\cal O}(N_c^{-1})$.
As an example, lets us consider 
\bey
\tr[\partial_\mu U\partial^\mu U^\dagger],
\label{kineticterm}
\eey
which appears in the coefficient of $K^\dagger K$ in \eq{lag0}.
For $B=1$, under the replacement
\ben
u\rightarrow AuA^\dagger
\een
with $A$ being a collective coordinate, \eq{kineticterm} reads
\ben
\tr[\partial_\mu u\partial^\mu u^\dagger]\rightarrow
\tr[A \partial_iuA^\dagger A\partial^iu^\dagger A^\dagger]+{\cal O}(N_c^{-1})
=\tr[\partial_iu\partial^iu^\dagger]+{\cal O}(N_c^{-1}).
\een
Therefore, for $B=1$, the soliton may be regarded as a static one at $N_c\rightarrow\infty$.
On the other hand, in the present $B=2$ case, since each Skyrmion is rotated independently,
\ben
u\rightarrow A_1u(1)A^\dagger_1 A_2u(2)A^\dagger_2.
\een
\eq{kineticterm} reads as follows,
\bey
\tr[\partial_\mu u\partial^\mu u^\dagger]&\rightarrow&
\tr[\partial_i(A_1u(1)A^\dagger_1 A_2u(2)A^\dagger_2)\partial^i(A_2u^\dagger(2)A^\dagger_2A_1u^\dagger(1)A^\dagger_1)]+{\cal O}(N_c^{-1})\cr
&=&\tr[ (A_1 \partial_i u(1)A^\dagger_1 A_2u(2)A^\dagger_2+A_1u(1)A^\dagger_1 A_2 \partial_i u(2)A^\dagger_2)\cr
&&\times
(A_2 \partial^i u^\dagger(2)A^\dagger_2A_1u^\dagger(1)A^\dagger_1+A_2u^\dagger(2)A^\dagger_2A_1 \partial^i u^\dagger(1)A^\dagger_1)] \cr
&=&\tr[\partial_i u(1)A^\dagger_1 A_2u(2) \partial^i u^\dagger(2)A^\dagger_2A_1u^\dagger(1)
+ \partial_i u(2) \partial^i u^\dagger(2) \cr
&&
+ \partial_i u(1)\partial^i u^\dagger(1)
+ u(1)A^\dagger_1 A_2 \partial_i u(2) u^\dagger(2)A^\dagger_2A_1 \partial^i u^
\dagger(1)]+{\cal O}(N_c^{-1}).
\eey
Thus the collective coordinates themselves, which are not suppressed by $1/N_c$, enters the Lagrangian.
As long as using the ansatz, \eq{ansatz}, the kaon inevitably couples not to static soliton but to rotating solitons in the limit of $N_c\rightarrow\infty$.
Accordingly, we consider the kaon under the background of the two-Skyrmion projected onto a spin-isospin state of two-nucleon.

Our next task is to perform the collective coordinate quantization, 
and project the rotation of each Skyrmion onto the relevant spin-isospin state.
This procedure is done as follows.
First, we rewrite the Lagrangian, \eq{lag}, in terms of the adjoint matrix defined by 
\ben
D_{ij}(A)=\tr(\tau_iA\tau_jA^\dagger)/2
\label{adjmat}
\een
with $A$ being a collective coordinate.
The result, which is quite lengthy, is displayed in Appendix \ref{lag-Dij}.
The matrix $D_{ij}(A)$ is known to be represented by the rotation matrix of rank-1.
On the other hand, the wave function of the nucleon in the space of collective coordinate is also expressed by rotation matrix \cite{adkins}:  
\ben
\langle A|N_{I_3,J_3}\rangle=\frac{1}{2\pi}(-1)^{I_3+1/2}D^{1/2}_{-I_3 J_3}(\Omega),
\label{nuclwf}
\een
where $I_3$ and $J_3$ denote the third component of the isospin and that of the spin, respectively, and $\Omega$ the Euler angles. 
Then, the projection of the Skyrmions onto physical two nucleon states
is performed by sandwiching the Lagrangian, \eq{lag}, with two nucleon states and integrating the Euler angles,
\bey
\int d\Omega_1\int d\Omega_2\langle N(1)N(2)|{\cal L}|N(1)N(2)\rangle,
\label{lagKNN}
\eey
where $N(i)$ denotes the i-th nucleon. 
We assume that the proton-proton in the $ppK^-$ system is in spin-singlet,
and project the rotational motion of the Skyrmions onto the spin-singlet proton-proton state.
Thus we consider
\bey
{\cal L}_{ppK}
\equiv\int d\Omega_1\int d\Omega_2\langle (pp)_{s=0}|{\cal L}|(pp)_{s=0}\rangle,
\label{lagKpp}
\eey
where $|(pp)_{s=0}\rangle$ is the wave function corresponding to the spin-singlet proton-proton state and is given by
\ben
|(pp)_{s=0}\rangle=
\frac{1}{\sqrt{{\cal N}}}(|N_{\frac{1}{2},\frac{1}{2}}(1)N_{\frac{1}{2},\frac{-1}{2}}(2)\rangle
-|N(1)_{\frac{1}{2},\frac{-1}{2}}N_{\frac{1}{2},\frac{1}{2}}(2)\rangle),
\een
with $\cal N$ being the normalization constant.
\eq{lagKpp} is the Lagrangian for the kaon coupled to two protons.
Detailed description of the projection are shown in Appendix \ref{formula-projection}.

Now, we derive the equation of motion for the kaon from the Lagrangian, \eq{lagKpp}.
First, we average the direction of the line joining the two Skyrmions.
To do that, we put $\vecR/2=((R/2)\sin\alpha\cos\beta, (R/2)\sin\alpha\sin\beta,(R/2)\cos\alpha)$ in the Lagrangian \eq{lagKpp}, and integrate the angles $\alpha$ and $\beta$,
\ben
\bar{\cal L}_{ppK}=\frac{1}{4\pi}\int_0^{\pi}d\alpha\int_0^{2\pi}d\beta\sin\alpha {\cal L}_{ppK}.
\label{avlag}
\een
This corresponds to assuming that the proton-proton system is in $S$-wave.
Then the background field becomes spherical, which allows us to set the kaon field as
\bey
K(\vecr,t)=k(r,t)Y_{lm}(\theta,\phi)
\label{kansatz}
\eey
with $Y_{lm}(\theta,\phi)$ the spherical harmonics.
This ansatz, \eq{kansatz}, is substituted into the Lagrangian, \eq{avlag}.
We perform the $\theta$- and $\phi$-integrations before taking the variation with respect to $k(r,t)$.
Up to this step, quite long and involved calculations are needed. 
Then the Euler-Lagrange equation for $k(r,t)$ yields
\bey
&&
\Big[-\bar{f}(r;R)\frac{d^2}{dt^2}-2i\bar{\lambda}(r;R)\frac{d}{dt}
-m_K^2-\bar{V}_{\rm eff}(r;R,l)+\hat{\cal O}\Big]k(r,t)=0,
\label{eom}
\eey
where the operator $\hat{\cal O}$ is defined as
\bey
\hat{\cal O}&=&
c_1(r;R)\frac{\partial}{\partial r}+c_2(r;R)\frac{\partial^2}{\partial r^2}.
\label{ohat}
\eey
In \eqs{eom}{ohat}, the coefficients, $\bar{f}(r;R)$ and $\bar{\lambda}(r;R)$ are expressed as follows:
\bey
{\bar f}(r;R)&=&\frac{1}{4\pi}\int_0^{\pi}d\alpha\int_0^{2\pi}d\beta\sin\alpha 
\left\langle1+{1\over2e^2F_K^2}\tr(\partial_j U_{BB}^\dagger\partial_j U_{BB})\right\rangle,\\
{\bar \lambda}(r;R)&=&\frac{1}{4\pi}\int_0^{\pi}d\alpha\int_0^{2\pi}d\beta\sin\alpha {-N_c\over F_K^2}\left\langle B^0\right\rangle.
\eey
Here $\langle\cdots\rangle$ means taking an expectation value with respect to the spin-singlet proton-proton state as in \eq{lagKpp}.
$-\bar{V}_{\rm eff}(r;R,l)$ and $\hat{\cal O}$ correspond to the terms with and without spatial derivative in the following equation, respectively,
\bey
-\bar{V}_{\rm eff}(r;R,l)+\hat{\cal O}&=&
\int_0^{\pi}d\theta\int_0^{2\pi}d\phi\sin\theta Y_{lm}(\theta,\phi)
\cdot\frac{1}{4\pi}\int_0^{\pi}d\alpha\int_0^{2\pi}d\beta\sin\alpha
\cr&&\times
\left\langle
D_jD_j-{1\over8}\left\{-\tr(\partial_j U_{BB}^\dagger\partial_j U_{BB})
+{1\over e^2F_K^2}\tr[\partial_j U_{BB}U_{BB}^\dagger,\partial_i U_{BB} U_{BB}^\dagger]^2\right\}\right.
\cr&&\left.
+{1\over e^2F_K^2}\left\{
2D_j[D_i \tr(A_j A_i)]
+{1\over2}D_j[D_j\tr(\partial_iU_{BB}^\dagger\partial_iU_{BB})]
-6D_j([A_i,A_j]D_i)\right\}
\right\rangle Y_{lm}(\theta,\phi).
\eey
$c_1$ and $c_2$ in $\hat{\cal O}$, and $\bar{V}_{\rm eff}(r;R,l)$ are read off from this equation. Their explicit expressions are quite lengthy and are not particularly instructive.
Therefore we do not display them here.

Let us expand the field $k(r,t)$ in terms of its eigenmodes:
\bey
k(r,t)&=&
\sum_n\left[k_n(r)e^{i\omega_nt}a_n^\dagger
+\tilde{k}_n(r)e^{-i\tilde{\omega}_nt}b_n\right],\quad (\omega_n,\,\tilde{\omega}_n>0),
\label{modeexpand}
\eey
where $a_n$ and $b_n$ are the annihilation operators for the strangeness $S=\mp 1$ states, respectively.
Substituting \eq{modeexpand} into \eq{eom}, we find the eigenmodes satisfy
\bey
&&\Big[\bar{f}(r;R)\omega_n^2-m_K^2-V_{K}^{(-)}(r;\omega_n,R,l)+\hat{\cal O}\Big]k_n(r)=0,\,\,
\label{eoms-1}\\
&&\Big[\bar{f}(r;R)\tilde{\omega}_n^2-m_K^2-V_{K}^{(+)}(r;\tilde{\omega}_n,R,l)+\hat{\cal O}\Big]\tilde{k}_n(r)=0.\,\,
\label{eoms+1}
\eey
\eqs{eoms-1}{eoms+1} are the equation of motions for $S=-1$ and $S=+1$ states, respectively.
In \eqs{eoms-1}{eoms+1}, $V_{K}^{(\mp)}(r;\omega,R,l)$ plays a role of potential term and can be separated into two terms,
\bey
V_{K}^{(\mp)}(r;\omega,R,l)=V_{WZW}^{(\mp)}(r;\omega,R)+\bar{V}_{\rm eff}(r;R,l),
\label{pot}
\eey
where $V_{WZW}^{(\mp)}(r;\omega,R)$ originates from the WZW term and $\bar{V}_{\rm eff}(r;R,l)$ from remaining terms in the Skyrme Lagrangian. $V_{WZW}^{(\mp)}(r;\omega,R)$ is given by
\bey
V_{WZW}^{(\mp)}(r;\omega,R)=\mp2\bar{\lambda}(r;R)\omega.
\label{potWZ}
\eey
Thus the WZW term acts on the negative (positive) strangeness states in
attarctive (repulsive) way.
$k_n(r)$ and $\tilde{k}_n(r)$ obey the following normalization conditions,
\bey
&&\int 4\pi r^2dr\left[\bar{f}(r;R)(\omega_n+\omega_{n'})+2\bar{\lambda}(r;R)\right]k_n(r)k_{n'}(r)=\delta_{nn'},
\label{normcond1}\\
&&\int 4\pi r^2dr\left[\bar{f}(r;R)(\tilde\omega_n+\tilde\omega_{n'})-2\bar{\lambda}(r;R)\right]\tilde k_n(r)\tilde k_{n'}(r)=\delta_{nn'}.
\label{normcond2}
\eey
\section{numerical solution of the kaon equation of motion}
\label{eomsol}
\begin{figure}[t]
\begin{center}
\rotatebox{-90}
{\includegraphics[width=6cm,keepaspectratio]{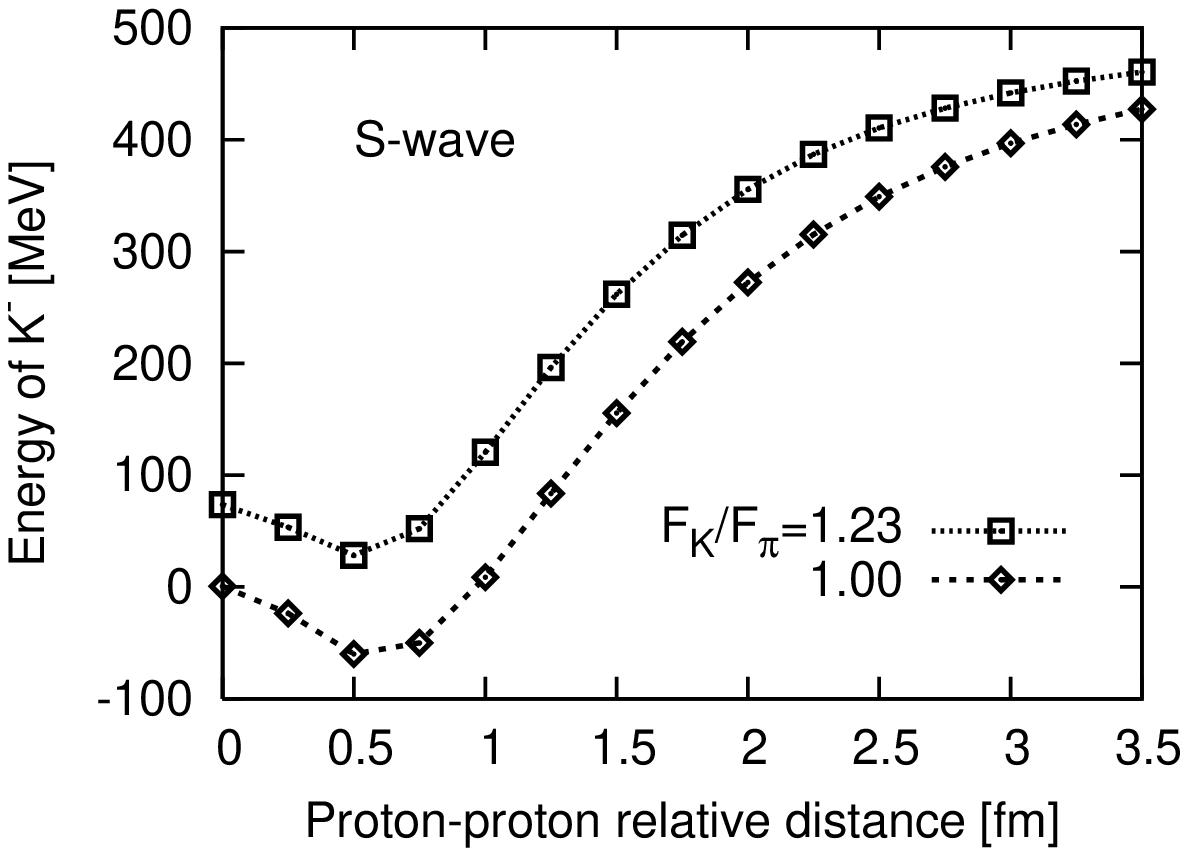}}
\rotatebox{-90}
{\includegraphics[width=6cm,keepaspectratio]{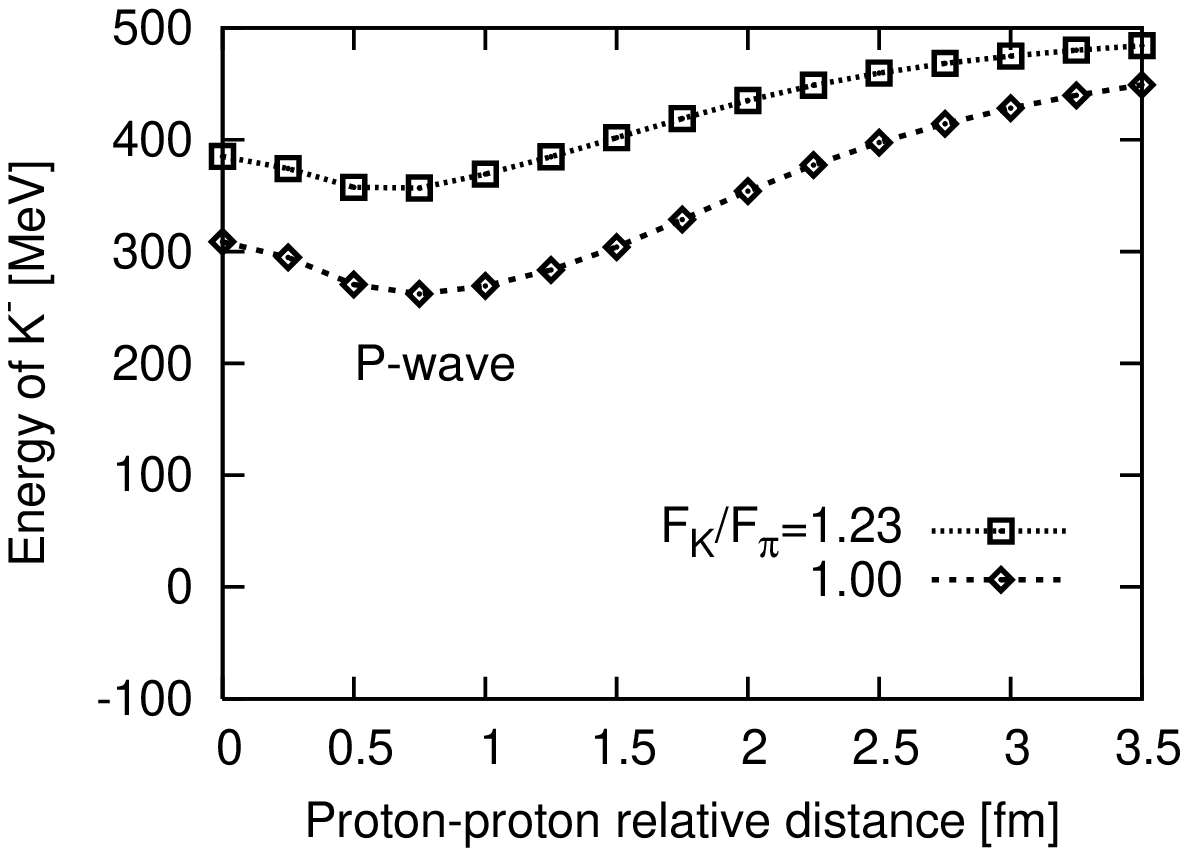}}
\caption{The energy eigenvalues of $S$- and $P$-wave $K^-$, $\omega$, as functions of the proton-proton relative distance, $R$. Two choices of the ratio $F_K/F_\pi$ are examined:
$F_K/F_\pi=1.00$ and the experimental value, $F_K/F_\pi=1.23$.}
\label{omega}
\end{center}
\end{figure}
\begin{table}
\begin{center}
\begin{tabular}{ccc}
\hline
\multicolumn{1}{c}{$R$ (fm)}&\multicolumn{1}{c}{$\omega_{l=0}$\,(MeV)} & \multicolumn{1}{c}{$\omega_{l=1}$\,(MeV)} \\
\hline\hline
\multicolumn{1}{c}{1.0}   & \multicolumn{1}{c}{121}  &\multicolumn{1}{c}{369}\\
 \multicolumn{1}{c}{1.5}  & \multicolumn{1}{c}{262}  & \multicolumn{1}{c}{402}\\
 \multicolumn{1}{c}{2.0}  & \multicolumn{1}{c}{356}  &\multicolumn{1}{c}{435}\\
  \multicolumn{1}{c}{2.5} & \multicolumn{1}{c}{411}  & \multicolumn{1}{c}{460}\\
  \multicolumn{1}{c}{3.0} & \multicolumn{1}{c}{442}  & \multicolumn{1}{c}{475}\\
  \hline
\end{tabular}
\end{center}
\caption{Energy eigenvalue of $S$-wave $K^-$ ($\omega_{l=0}$) and that of $P$-wave ($\omega_{l=1}$) for five cases of the proton-proton relative distance, $R$.
$F_K/F_\pi$ is taken to be the empirical value, $F_K/F_\pi=1.23$.}
\label{tableomega}
\end{table}
\begin{figure}[t]
\begin{center}
\rotatebox{-90}
{\includegraphics[width=6cm,keepaspectratio]{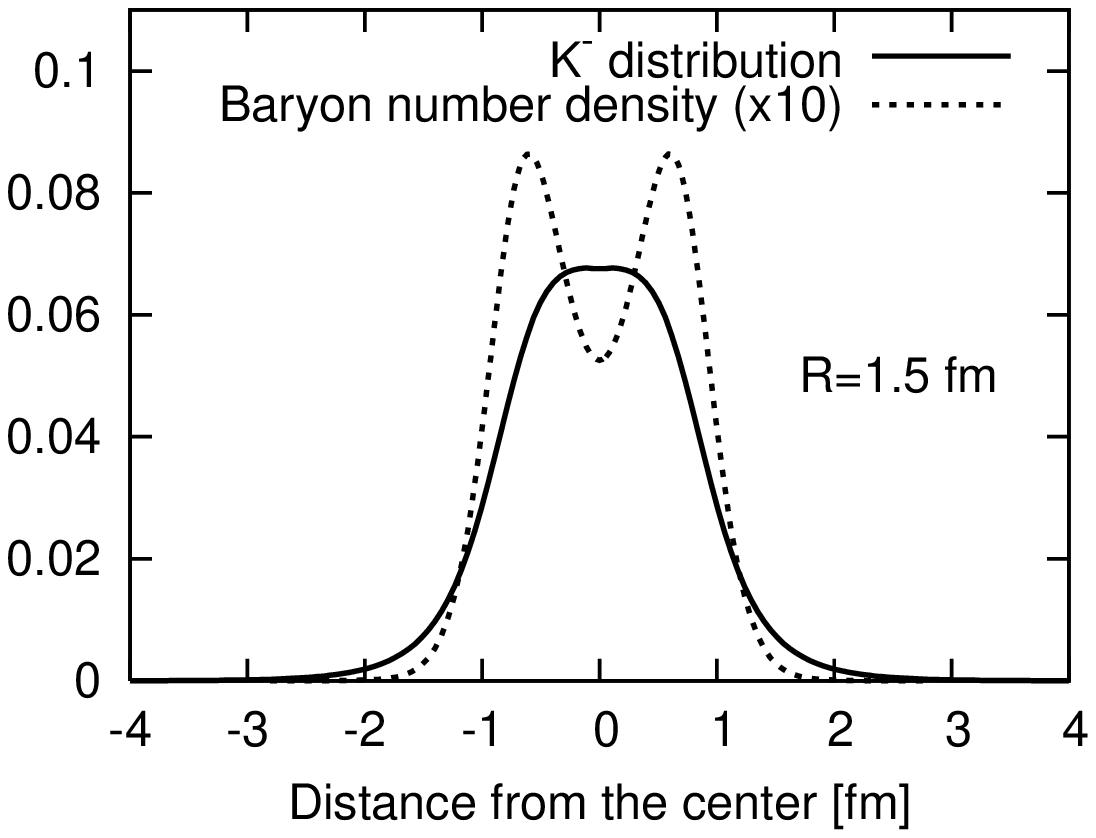}}
\rotatebox{-90}
{\includegraphics[width=6cm,keepaspectratio]{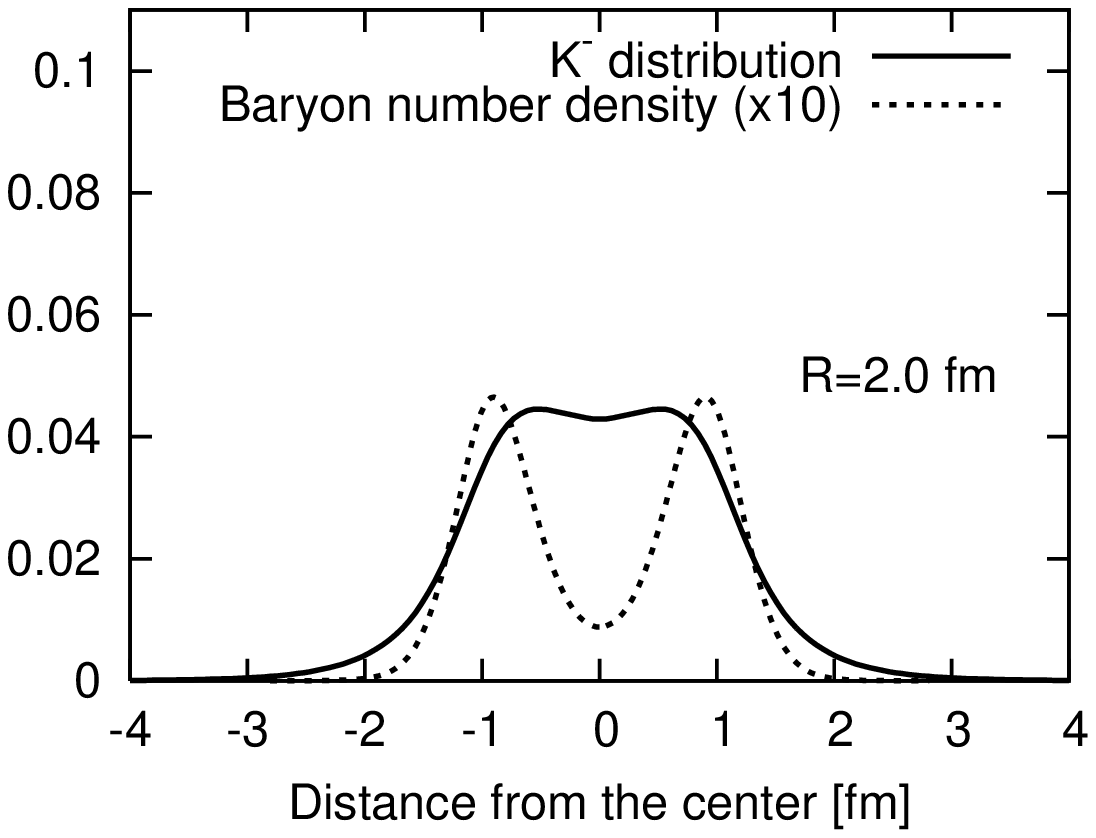}}
\rotatebox{-90}
{\includegraphics[width=6cm,keepaspectratio]{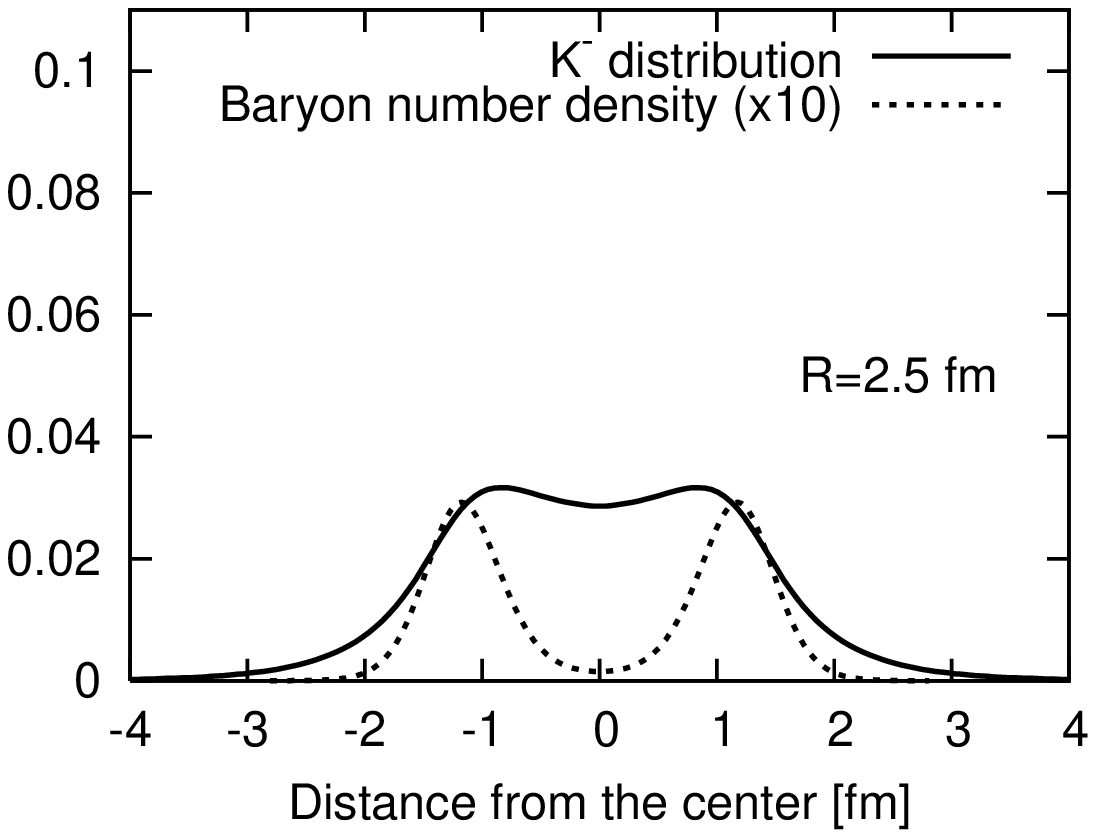}}
\rotatebox{-90}
{\includegraphics[width=6cm,keepaspectratio]{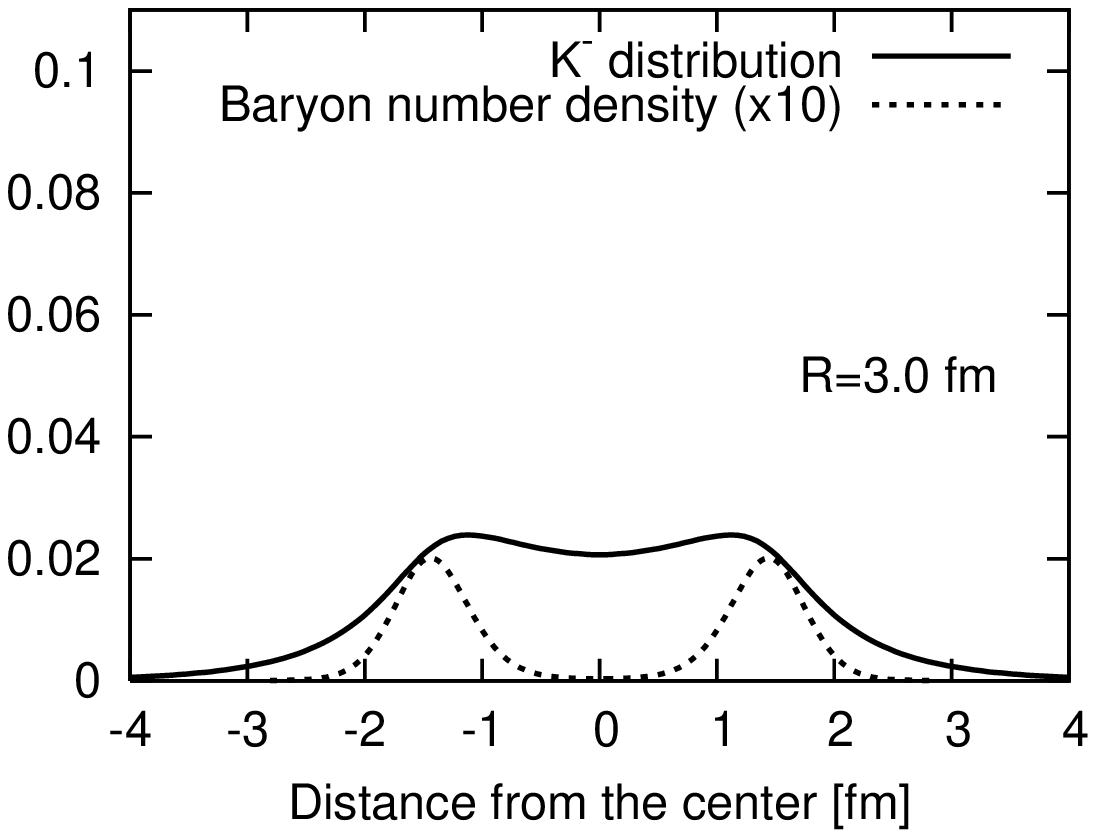}}
\caption{$S$-wave $K^-$ distribution (normalized wave function $k(r)$) and baryon number density (\eq{bpp}) for the relative distance of the two Syrmions, $R=1.5$, $2.0$, $2.5$ and $3.0\,{\rm fm}$. Both are in unit of $eF_\pi$. The horizontal axis is the distance from the origin. $F_K/F_\pi=1.23$ was chosen.
The baryon number density is multiplied by a factor $10$.}
\label{wfs}
\end{center}
\end{figure}

We solved numerically the equation of motion, \eq{eoms-1}.
Figure.\ref{omega} shows the obtained energy eigenvalue of $K^-$, $\omega$, as a function of the Skyrmion-Skyrmion relative distance, $R$.
$m_\pi$, $F_\pi$ and $e$ were taken to be $m_\pi=0$, $F_\pi=129$ MeV and $e=5.45$, which were adjusted to fit the masses of $N$ and $\Delta$ \cite{adkins}. The kaon mass was taken to be $m_K=495$ MeV. For the ratio, $F_K/F_\pi$, we have examined two choices:
$F_K/F_\pi=1.00$ \footnote{In our previous paper \cite{NK}, we adopted this choice.} and the empirical value, $F_K/F_\pi=1.23$. 

We find that the lowest-lying mode is the $S$-wave and that the $P$-wave lies above the $S$-wave.
Note that this order is natural but different from the case of $B=1$, where
the lowest-lying mode is $P$-wave, as mentioned in the introduction.
We see that it is important to take into account the difference between $F_K$ and $F_\pi$. As was shown in Ref.\cite{sco}, by setting $F_K/F_\pi$ equal to the empirical vaue, $F_K/F_\pi=1.23$, hyperon masse are well reproduced, while when we set $F_K/F_\pi=1$ they are overbound.
The binding of $K^-$ to $pp$ is also weaker when taking the empirical value of $F_K/F_\pi$.   

In Table \ref{tableomega}, the $K^-$ energy eigenvalues for several values of $R$ are displayed. 
Looking at the $S$-wave channel, the binding of the kaon is extremely strong
for smaller distance, {\it i.e.}  $R\aplt 1.0\,{\rm fm}$. 
In this region, the repulsive nucleon-nucleon interaction dominates over the attractive interaction between $K^-$ and $pp$.
Nuclear matter in which average $NN$ distance is $R\simeq 1.0$ fm is expected to be realized in the core region of compact stars.
Our result shows that $K^-$ in such high density nuclear matter can
be lighter than the pion.
This might be somehow related with the kaon condensation \cite{kaplan} which is considered to occur in compact stars.
As $R$ is increased, the binding becomes looser.
However, at $R=2.0\,{\rm fm}$, for instance, which is close to the average inter $NN$ distance in normal nuclei, the binding is still deep: the binding energy is about $140\,{\rm MeV}$ (see Table \ref{tableomega}).  

Next, we consider the $R$-dependence of the $K^-$-distribution.
In Figure \ref{wfs},
we plot the distribution of $K^-$ in $S$-wave (kaon wave function $k(r)$ normalized by the condition, \eq{normcond1}) for several cases of $R$, $R=1.5$, $2.0$, $2.5$ and $3.0\,{\rm fm}$.
For comparison, the baryon number denisity given by
\bey
\langle B^0\rangle\equiv\int d\Omega_1\int d\Omega_2\langle (pp)_{s=0}|B^0|(pp)_{s=0}\rangle,
\label{bpp}
\eey
is also plotted.
At $R=1.5\,{\rm fm}$, it can be seen that $K^-$ is localized in the narrow region between the two protons.
At relatively larger separation, $R\aplg 2.0\,{\rm fm}$, $K^-$ has large probability to stay near the proton's respective positions, which is characteristic to molecular orbital states \cite{NK}.

\section{proton-proton radial motion}
\begin{figure}[t]
\begin{center}
\rotatebox{-90}
{\includegraphics[width=6cm,keepaspectratio]{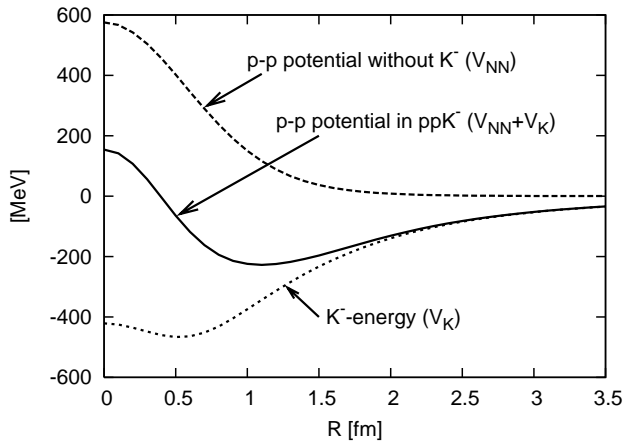}}
\caption{Potential terms in \eq{hamiltonian} as functions of the proton-proton relative distance, $R$. The upper curve is the proton-proton potential in the absence of $K^-$, $V_{NN}(R)$. The lower one represents the energy of $K^-$, $V_{K}(R)$. The middle one corresponds to their sum, $V_{NN}(R)+V_{K}(R)$, the effective proton-proton potential in the $ppK^-$ system.}
\label{potential}
\end{center}
\end{figure}

Now, we solve the dynamical problem of proton-proton radial motion under the strong attaractive potential mediated by $K^-$, in addition to the ordinary nucleon-nucleon potential.
We assume that the radial motion of the two-proton is governed by the following Hamiltonian,
\bey
H=T_{NN}(R)+V_{NN}(R)+V_{K}(R),
\label{hamiltonian}
\eey
where $T_{NN}$ is the $NN$ kinetic energy term,
\bey
T_{NN}=-\frac{1}{M_N}\frac{1}{R^2}\frac{\partial}{\partial R}\left(R^2\frac{\partial}{\partial R}\right).
\eey
Here, the nucleon is regarded as not a soliton with finite size but a point like particle with the mass $M_N=939$ MeV and its motion is assumed to be non relativistic.
$V_{NN}(R)$ is the state-independent part of the $NN$ potential obtained from the product of $B=1$ Skyrmion \cite{yabu}.
We have neglected the contribution from the state-dependent part since they give a minor contribution compared with the state-independent part.
The last term is the energy generated by the bound kaon in $S$-wave, 
\bey
V_{K}(R)=\omega_{l=0}(R)-m_K,
\eey
where $\omega_{l=0}(R)$ is the $S$-wave kaon's energy obtained in the previous section.  
$V_{NN}(R)+V_{K}(R)$ may be regarded as effective $pp$ potential in the $ppK^-$ system.
In the left panel of Figure \ref{potential}, we show the behavior of $V_{NN}(R)$, $V_{K}(R)$ and their sum.
$V_{NN}(R)$ has repulsive core at short distances and one-pion exchange tail at long distances. Medium range attraction cannot be produced within the product ansatz of $B=1$ Skyrmions \cite{oka}.
On the other hand,  the attractive force generated by bound kaon, $V_{K}(R)$,  is so strong that it overcomes the strongly repulsive $V_{NN}(R)$.
As a result, the effective $NN$ potential in the $ppK^-$ system, $V_{NN}(R)+V_{K}(R)$, is strongly attractive in the medium range. 

The energy of the $ppK^-$ state relative to $2M_N+m_K$, $E$, is obtained by solving the Schrodinger equation,
\bey
H\Psi_N=E\Psi_N.
\label{sch}
\eey
In Table \ref{be}, we show the energy of $ppK^-$ bound state relative to $2M_N+m_K$ and its decomposition into the $NN$ kinetic energy, $\langle T_{NN}\rangle$, and the $NN$ potential energy, $\langle V_{NN}\rangle$, and the kaon's energy, $\langle V_{K}\rangle$, obtained by solving \eq{sch}.
The expectation value, $\langle{\cal O}\rangle$, is defined by
\bey
\langle {\cal O}\rangle\equiv\int_0^\infty 4\pi R^2dR \Psi_N(R)^*{\cal O}\Psi_N(R)\Big/\int_0^\infty 4\pi R^2dR \Psi_N(R)^*\Psi_N(R).
\eey
The root mean square radius for $NN$ subsystem, $\sqrt{\langle r_{NN}^2\rangle}$, is also shown.

When the $F_K/F_\pi$ is taken to be the experimental value, $F_K/F_\pi=1.23$, 
the binding energy of the $ppK^-$ bound state is estimated to be
\bey
B_{ppK^-}\simeq 126\,{\rm MeV}.
\eey
Another fact worth noting is the smallness of the $NN$ kinetic energy,
$\langle T_{NN}\rangle\sim 40$ MeV. As long as looking at this fact, the Born-Oppenheimer approximation seems to be not so poor.
The root mean square radius of the two-nucleon subsystem is 
\bey
\sqrt{\langle r_{NN}^2\rangle}\simeq 1.6\,{\rm fm},
\eey
which is significantly smaller than the average inter $NN$ distance in normal nuclei.

\begin{table}[t]
\begin{center}
\begin{tabular}{cccccc}
\hline
$F_K/F_\pi$ & $\langle T_{NN}\rangle$ (MeV) & $\langle V_{NN}\rangle$  (MeV)& $\langle V_{K}\rangle$  (MeV) & total (MeV)& $\sqrt{\langle r_{NN}^2\rangle}$ (fm)\\
\hline\hline
1.00 & 50.4 & 97.3 & -380.5 & -232.7 & 1.46\\
\hline
1.23 & 42.0 & 74.5 & -239.2 & -125.5 & 1.63\\
\hline
\end{tabular}
\end{center}
\caption{Calculated total energy of the $ppK^-$ bound state relative to $2M_N+m_K$ and its decomposition. The root mean square radius of $NN$ subsystem,  $\sqrt{\langle r_{NN}^2\rangle}$, is also shown. For the definition of each component, see the text.}
\label{be}
\end{table}%

\section{discussion and summary}
\begin{figure}[t]
\begin{center}
\rotatebox{-90}
{\includegraphics[width=6cm,keepaspectratio]{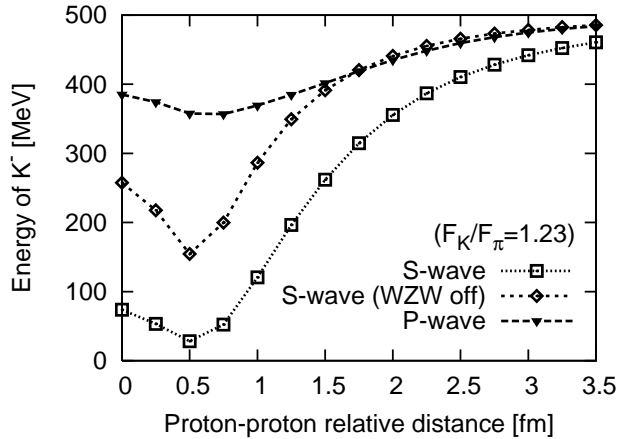}}
\caption{The energy eigenvalue of $K^-$ as a function of the proton-proton relative distance, $R$, compared with the case when the Wess-Zumino-Witten term is switched off. The ratio of the decay constant is taken to be $F_K/F_\pi=1.23$. 
The bottom and upper curves are the energies of $K^-$ in $S$- and $P$- wave, respectively. The middle one represents the energy of $S$-wave $K^-$ when the Wess-Zumino-Witten term is switched off.
$P$-wave $K^-$ is unbound when the Wess-Zumino-Witten term is switched off.
}
\label{omega-wzoff}
\end{center}
\end{figure}

Lets us consider the mechanism responsible for the strong binding of $K^-$ to $pp$.
This is turned out to be attributed to the WZW term.
It is known that in $B=1$ sector the existence of the WZW term leads to various important results.
The WZW term is included in the action from the requirement that the effective theory written in terms of the Nambu-Goldstone bosons should reproduce the anomaly the fundamental theory possesses.
The Skyrme Lagrangian without the term has a fictitious symmetry: an invariance under $U\leftrightarrow U^\dagger$.
This symmetry forbids processes changing even-oddness of meson number, {\it e.g.} $K^+K^-\rightarrow\pi^+\pi^-\pi^0$. 
The WZW term breaks this extra symmetry \cite{witten2}. 
In addition, for odd $N_c$, the WZW term ensures that the quantized Skyrmion has a half-odd spin and thus behaves as a fermion.  
The effect of the WZW term goes beyond these rules. 
In the bound kaon approach, the interaction Lagrangian of the kaon and the nucleon which comes from the WZW term has the form of
\bey
{\cal L}_{\rm WZW}=-{iN_c\over F_K^2}[K^\dagger \partial_0K
-(\partial_0K)^\dagger K]B^0.
\label{LWZW}
\eey
This is nothing but the so called Tomozawa-Weinberg term.
This term gives an effective attractive contribution to negative strangeness states which is crucial for obtaining the correct values of the masses of ground state hyperons.
In particular, without the term, the $S$-wave bound state of a kaon and a Skyrmion, which corresponds to $\Lambda(1405)$, does not exist.
Also in the present $B=2$ case, the role of the WZW term is revealed to be important.
In the equation of motion, \eq{eoms-1}, there exist two terms, $V_{WZW}^{(-)}$ and $\bar{V}_{\rm eff}$ in \eq{pot}, which effectively play a role of the potential acting on the kaon. Among them, it is $V_{WZW}^{(-)}$ that originates from the WZW term.
In order to see the effects of the WZW term, we switched off $V_{WZW}^{(-)}$ and calculated the kaon's energy.
The result for the $S$-wave $K^-$ is shown in Figure \ref{omega-wzoff}.
One observes that the WZW term additionally gives a substantial attractive contribution to the binding of the kaon.
Here, we should note that $V_{WZW}^{(-)}$ is stronger than that for $B=1$ since the interaction \eq{LWZW} is proportional to the baryon number density.
It is therefore quite a natural result that $K^-$ is bound to two-proton more deeply than to one proton.
However, this does not necessarily mean that the larger the baryon number $B$ becomes the deeper the binding of the kaon is,
since $V_{WZW}^{(-)}$ does not necessarily become stronger with the increasing  $B$: $V_{WZW}^{(-)}$ is proportional to the kaon's energy (see \eq{potWZ}).

The $K^-$-distribution shown in Figure \ref{wfs} suggests that 
the $ppK^-$ state is a molecular orbital state.
This observation is quite natural in the following sense.
The two protons in the $ppK^-$ system should keep some distance so as to avoid the repulsive core of the nuclear potential.
On the other hand, for the two-proton with finite separation, it can be shown that the potential acting on the kaon, $V_{K}^{(-)}(r;\omega,R,l)$ in \eq{pot}, is a double-well potential which is most attractive at the proton's respective position.
It is the molecular orbit that the kaon occupies under such a double-well potential.
If the $ppK^-$ is really a molecular orbital state, it is plausible that the binding of $K^-$ to two-proton is stronger than to one proton
since $K^-$ experiences the strong attraction from the two protons without increase of the kinetic energy. 

Finally, an important comment is in order. As was discussed in the section \ref{deriveom}, the kaon is inevitably bound to rotating solitons in the limit of $N_c\rightarrow\infty$, as far as we employ the product ansatz, \eq{ansatz}, and rotate the solitons independently. Therefore we have projected the Skyrmion rotation onto a relevant spin-isospin state of the two-nucleon, spin-singlet proton-proton state.
Thus the system which we have considered is purely a $ppK^-$ bound state.
On the other hand, in the original bound kaon approach, hyperons do not necessarily correspond to $\bar K N$ bound states but quantized states of the bound system of a $\rm SU(2)_f$ soliton and kaon fluctuating around the soliton.
It would be appropriate to investigate the $ppK^-$ states
by extending such a picture of the hyperon to $B=2$ and $S=-1$ systems,
which is left for our future work.

In summary, we have applied the Skyrme model to a study of the lightest $\bar{K}$-nuclear bound state, $ppK^-$.
We have derived the equation of motion for the kaon coupled to two-proton at fixed position.
The two-proton is expressed by two-Skyrmion whose rotational motion in the space of collective coordinate is quantized and projected onto spin-singlet proton-proton state.
Numerical solution of the equation of motion shows that $K^-$ can be strongly bound even for relatively large inter proton-proton distances and that $K^-$ is in a molecular orbital state.
Next, we have solved the two-proton radial motion by assuming that the protons move under the strong attractive potential generated by $K^-$ in addition to the ordinary $NN$ potential. 
Then we have found that $ppK^-$ state can be realized as a very deeply bound and compact state, whose binding energy is $B_{ppK^-}\simeq 126$ MeV and the mean inter $NN$ distance is $\sqrt{\langle r_{NN}^2\rangle}\simeq 1.6$ fm.
The obtained value of the binding energy is surprisingly close to the experimental result obtained by FINUDA collaboraton, $B_{ppK^-}^{\rm exp.}=115$ MeV.
However, considering the crudeness of our treatment, we are not allowed to satisfy this agreement.

\begin{acknowledgments}
This work was supported in part by the 21st Century COE Program at Tokyo
Institute of Technology ``Nanometer-Scale Quantum Physics'' from the
Ministry of Education, Culture, Sports, Science and Technology, Japan.
\end{acknowledgments}  
\appendix
\section{Full expression of the kaon Lagrangian}
\label{lag-Dij}
In this appendix, we rewrite the Lgrangian, \eq{lag}, in terms of the adjoint matrix, \eq{adjmat}.
The adjoint marix is included in $U_{BB}$, $V_j$ and $A_j$.
$V_j$ and $A_j$ can be rewritten using the adjoint matrix as follows,
\bey
V_j&=&i\tau_m{1\over2}[D_{mi}(A_1)L_j^i(1)+D_{mi}(A_2)R_j^i(2)],\\
A_j&=&i\tau_m{1\over2}[D_{mi}(A_1)L_j^i(1)-D_{mi}(A_2)R_j^i(2)].
\end{eqnarray}
Here $L_j^i(1)$ and $R_j^i(2)$ are defined as
\begin{eqnarray}
L_j^i(1)&=&\Ld_{ji}{1\over r(1)}C(1)S(1)+\Lr_j\Lr_iF'(1)
+\epsilon_{jik}\Lr_k{1\over r(1)}S(1)^2,\\
R_j^i(2)&=&-\Rd_{ji}{1\over r(2)}C(2)S(2)-\Rr_j\Rr_iF'(2)
+\epsilon_{jik}\Rr_k{1\over r(2)}S(2)^2,
\eey
where
\bey
&&\hat\delta(\alpha)_{ji}=\delta_{ji}-\hat r(\alpha)_j\hat r(\alpha)_i,\\
&&F(\alpha)=F(r(\alpha)),\quad F'(\alpha)=\frac{d F(\alpha)}{d r(\alpha)},\\
&&C(\alpha)=\cos F(\alpha),\quad S(\alpha)=\sin F(\alpha),\quad(\alpha=1,2).
\eey
Now we use these equations to rewrite the Lagrangian in terms of the adjoint matrix and Skyrmion profile function.
Let us show the result for each term in \eq{lag}. 
\subsection{$(\partial_0K)^\dagger \partial_0K$-term}
$\tr(\partial_j U_{BB}^\dagger\partial_j U_{BB})$ in the first term of \eq{lag}
is expressed as follows,
\begin{eqnarray}
\tr(\partial_j U_{BB}^\dagger\partial_j U_{BB})
=2\left[L_j^i(1)L_j^i(1)+R_j^i(2)R_j^i(2)-D_{ik}(A^\dagger_1A_2)2L_j^i(1)R_j^k(2)\right].
\label{tder}
\end{eqnarray}
\subsection{$K^\dagger K$-term}
The first term in the curly bracket of the fourth term is given in \eq{tder}. The second term in the curly bracket is rewritten as follows,
\begin{eqnarray}
&&{1\over e^2F_\pi^2}\tr[\partial_\mu U_{BB}U_{BB}^\dagger,\partial_\nu U_{BB} U_{BB}^\dagger]^2\cr
&=&{4\over e^2F_\pi^2}\{
[(L_j^k(1)L_i^k(1)+R_j^k(2)R_i^k(2))(L_j^l(1)L_i^l(1)+R_j^l(2)R_i^l(2))
\cr&&-(L_j^i(1)L_j^i(1)+R_j^i(2)R_j^i(2))^2]
\cr&&+D_{kn}(A^\dagger_1A_2)[-2(L_j^k(1)R_i^n(2)+L_i^k(1)R_j^n(2))(L_i^l(1)L_j^l(1)+R_i^l(2)R_j^l(2))
\cr&&+4L_j^k(1)R_j^n(2)(L_m^l(1)L_m^l(1)+R_m^l(2)R_m^l(2))]
\cr&&+D_{kn}(A^\dagger_1A_2)D_{lm}(A^\dagger_1A_2)[(L_j^k(1)R_i^n(2)+L_i^k(1)R_j^n(2))
(L_i^l(1)R_j^m(2)+L_j^l(1)R_i^m(2))
\cr&&-4L_j^k(1)R_j^n(2)L_i^l(1)R_i^m(2)]\}.
\end{eqnarray}
\subsection{Terms containing $D_jK$}
In this subsection, we display the results for the terms containing spatial ``covariant derivative", $D_jK$.
The second term in \eq{lag} is rewritten as
\begin{eqnarray}
&&K^\dagger D_jD_jK
=K^\dagger\{
[\partial_j^2-{1\over4}(L_j^i(1)L_j^i(1)+R_j^i(2)R_j^i(2))]
+i\tau_mD_{mi}(A_1){1\over2}[\partial_jL_j^i(1)+L_j^i(1)\partial_j]
\cr&&\qquad\qquad\qquad+i\tau_mD_{mi}(A_2){1\over2}[\partial_jR_j^i(2)+R_j^i(2)\partial_j]
+D_{ik}(A^\dagger_1A_2)[-{1\over2}L_j^i(1)R_j^k(2)]
\}K.
\eey
The three terms in the curly bracket of the fifth term have very lengthy expressions.\\
The first term in the bracket:
\bey
&&2K^\dagger D_j[D_i K\tr(A_j A_i)]\cr
&&=2K^\dagger\{
[-{1\over2}\partial_j(L_j^l(1)L_i^l(1)+R_j^l(2)R_i^l(2))\partial_i
\cr&&\qquad+{1\over4}(L_j^p(1)L_i^p(1)+R_j^p(2)R_i^p(2))(L_j^l(1)L_i^l(1)+R_j^l(2)R_i^l(2))]
\cr&&\quad+D_{lk}(A^\dagger_1A_2){1\over2}\partial_j(L_j^l(1)R_i^k(2)+L_i^l(1)R_j^k(2))\partial_i
\cr&&\quad-D_{pm}(A^\dagger_1A_2)D_{lk}(A^\dagger_1A_2){1\over4}
(L_j^p(1)R_i^m(2)+L_i^p(1)R_j^m(2))(L_j^l(1)R_i^k(2)+L_i^l(1)R_j^k(2))
\cr&&\quad-i\tau_nD_{nq}(A_1){1\over4}
[\partial_jL_i^q(1)(L_j^l(1)L_i^l(1)+R_j^l(2)R_i^l(2))+(L_j^l(1)L_i^l(1)+R_j^l(2)R_i^l(2))L_j^q(1)\partial_i]
\cr&&\quad-i\tau_nD_{nq}(A_2){1\over4}
[\partial_jR_i^q(2)(L_j^l(1)L_i^l(1)+R_j^l(2)R_i^l(2))+(L_j^l(1)L_i^l(1)+R_j^l(2)R_i^l(2))R_j^q(2)\partial_i]
\cr&&\quad+i\tau_nD_{nq}(A_1)D_{lk}(A^\dagger_1A_2){1\over4}
[\partial_jL_i^q(1)(L_j^l(1)R_i^k(2)+L_i^l(1)R_j^k(2))
\cr&&\quad\qquad+(L_j^l(1)R_i^k(2)+L_i^l(1)R_j^k(2))L_j^q(1)\partial_i]
\cr&&\quad+i\tau_nD_{nq}(A_2)D_{lk}(A^\dagger_1A_2){1\over4}
[\partial_jR_i^q(2)(L_j^l(1)R_i^k(2)+L_i^l(1)R_j^k(2))
\cr&&\quad\qquad+(L_j^l(1)R_i^k(2)+L_i^l(1)R_j^k(2))R_j^q(2)\partial_i]\}K.
\eey
The second term:
\bey
&&
{1\over2}K^\dagger D_j[D_jK\tr(\partial_iU_{BB}^\dagger\partial_iU_{BB})]
\cr&&=K^\dagger\{
[\partial_j(L_m^n(1)L_m^n(1)+R_m^n(2)R_m^n(2))\partial_j-{1\over4}(L_j^i(1)L_j^i(1)+R_j^i(2)R_j^i(2))^2]
\cr&&\quad-D_{nk}(A^\dagger_1A_2)2\partial_jL_m^n(1)R_m^k(2)\partial_j
\cr&&\quad+D_{il}(A^\dagger_1A_2)D_{nk}(A^\dagger_1A_2)L_j^i(1)R_j^l(2)L_m^n(1)R_m^k(2)
\cr&&\quad+i\tau_lD_{li}(A_1){1\over2}
[L_j^i(1)(L_m^n(1)L_m^n(1)+R_m^n(2)R_m^n(2))\partial_j
\cr&&\quad\qquad+\partial_j(L_m^n(1)L_m^n(1)+R_m^n(2)R_m^n(2))L_j^i(1)]
\cr&&\quad+i\tau_lD_{li}(A_2){1\over2}
[R_j^i(2)(L_m^n(1)L_m^n(1)+R_m^n(2)R_m^n(2))\partial_j
\cr&&\quad\qquad+\partial_j(L_m^n(1)L_m^n(1)+R_m^n(2)R_m^n(2))R_j^i(2)]
\cr&&\quad+i\tau_lD_{li}(A_1)D_{nk}(A^\dagger_1A_2)
2[-L_j^i(1)L_m^n(1)R_m^k(2)\partial_j-\partial_jL_m^n(1)R_m^k(2)L_j^i(1)]
\cr&&\quad+i\tau_lD_{li}(A_2)D_{nk}(A^\dagger_1A_2)
2[-R_j^i(2)L_m^n(1)R_m^k(2)\partial_j-\partial_jL_m^n(1)R_m^k(2)R_j^i(2)]
\}K.
\eey
The third term: 
\bey
&&
-6K^\dagger D_j[A_i,A_j]D_iK\cr
&&=-3K^\dagger\{
-{1\over4}[(L_j^l(1)L_j^l(1)-R_j^l(2)R_j^l(2))(L_i^k(1)L_i^k(1)-R_i^k(2)R_i^k(2))
\cr&&\quad\qquad-(L_j^k(1)L_i^k(1)-R_j^k(2)R_i^k(2))(L_j^l(1)L_i^l(1)-R_j^l(2)R_i^l(2))]
\cr&&\quad
-i\tau_mD_{mk}(A_1)[\epsilon_{knl}\partial_jL_i^n(1)L_j^l(1)\partial_i
\cr&&\quad
\qquad-{1\over2}[L_i^k(1)(L_j^l(1)L_j^l(1)-R_j^l(2)R_j^l(2))
-L_j^k(1)(L_i^l(1)L_j^l(1)-R_i^l(2)R_j^l(2))]\partial_i
\cr&&\quad
\qquad+{1\over2}\partial_j[L_i^k(1)(L_i^l(1)L_j^l(1)-R_i^l(2)R_j^l(2))
-L_j^k(1)(L_i^l(1)L_i^l(1)-R_i^l(2)R_i^l(2))]]
\cr&&\quad
-i\tau_mD_{mk}(A_2)[\epsilon_{knl}\partial_jR_i^n(2)R_j^l(2)\partial_i
\cr&&\quad
\qquad+{1\over2}[R_i^k(2)(L_j^l(1)L_j^l(1)-R_j^l(2)R_j^l(2))
-R_j^k(2)(L_i^l(1)L_j^l(1)-R_i^l(2)R_j^l(2))]\partial_i
\cr&&\quad
\qquad-{1\over2}\partial_j[R_i^k(2)(L_i^l(1)L_j^l(1)-R_i^l(2)R_j^l(2))
-R_j^k(2)(L_i^l(1)L_i^l(1)-R_i^l(2)R_i^l(2))]]
\cr&&\quad
+i\tau_mD_{mk}(A_1)D_{lr}(A^\dagger_1A_2)[\epsilon_{kln}\partial_jL_j^n(1)R_i^r(2)\partial_i
\cr&&\quad
\qquad-{1\over2}L_j^k(1)(L_i^l(1)R_j^r(2)-L_j^l(1)R_i^r(2))\partial_i
-{1\over2}\partial_jL_i^k(1)(L_j^l(1)R_i^r(2)-L_i^l(1)R_j^r(2))]
\cr&&\quad
+i\tau_mD_{mk}(A_2)D_{lr}(A^\dagger_1A_2)[\epsilon_{krn}\partial_jL_i^l(1)R_j^n(2)\partial_i
\cr&&\quad
\qquad+{1\over2}R_j^k(2)(L_i^l(1)R_j^r(2)-L_j^l(1)R_i^r(2))\partial_i
+{1\over2}\partial_jR_i^k(2)(L_j^l(1)R_i^r(2)-L_i^l(1)R_j^r(2))]
\cr&&\quad
+D_{pr}(A^\dagger_1A_2)
[\epsilon_{klp}L_i^k(1)L_j^l(1)R_j^r(2)\partial_i+\epsilon_{klr}L_j^p(1)R_i^k(2)R_j^l(2)\partial_i
\cr&&\quad
\qquad+\epsilon_{klp}\partial_jL_i^k(1)L_j^l(1)R_i^r(2)+\epsilon_{klr}\partial_jL_i^p(1)R_i^k(2)R_j^l(2)]
\cr&&\quad+D_{kr}(A^\dagger_1A_2)D_{lt}(A^\dagger_1A_2)
{1\over4}(L_i^k(1)R_j^r(2)-L_j^k(1)R_i^r(2))(L_j^l(1)R_i^t(2)-L_i^l(1)R_j^t(2))
\}K.
\end{eqnarray}
\subsection{WZW-term}
The last term in \eq{lag}, which is proportional to the baryon number density, $B^0$, comes from the WZW term.
$B^0$ can be written in terms of $L$ and $R$ as
\begin{eqnarray}
&&B^0=-{1\over12\pi^2}
[\epsilon^{ijk}\epsilon_{pqr}(L_i^p(1)L_j^q(1)L_k^r(1)-R_i^p(2)R_j^q(2)R_k^r(2))
\cr&&\quad\quad
+D_{ps}(A^\dagger_1A_2)
3\epsilon^{ijk}(\epsilon_{qrs}L_i^p(1)R_j^q(2)R_k^r(2)-\epsilon_{qrp}R_i^s(2)L_j^q(1)L_k^r(1))].
\end{eqnarray}
\section{Spin-isospin projection}
\label{formula-projection}
In this appendix, we show the procedure of spin-isospin projection in detail.
The first step is to replace the adjoint matrix in the Lagrangian with Wigner $D$-function. For example,
\bey
&&D_{mi}(A_1)\rightarrow D^1_{MM'_i}(\Omega_1),\\
&&D_{ij}(A^\dagger_1A_2)=D_{mi}(A_1)D_{mj}(A_2)\rightarrow D^1_{MM'_i}(\Omega_1)D^1_{MM'_j}(\Omega_2).
\eey
Next, we sandwich them between the relevant nucleon wave function, \eq{nuclwf},
and integrate the Euler angle.
We demonstrate the procedure by taking two examples below.
For later use, we display two basic formulae for $D$-functions,
\begin{eqnarray}
&&D^{J_1}_{M_1M'_1}(\Omega)D^{J_2}_{M_2M'_2}(\Omega)
=\sum_{J=|J_1-J_2|}^{J_1+J_2}\<>{J_1J_2M_1M_2|JM}\<>{J_1J_2M'_1M'_2|JM'}D^{J}_{MM'}(\Omega),
\label{Dformula1}\\
&&\int d\Omega D^{J_1*}_{M_1M_1'}(\Omega)D^{J_2}_{M_2M_2'}(\Omega)
={8\pi^2\over2J_1+1}\delta(J_1,J_2)\delta(M_1,M_2)\delta(M'_1,M'_2).
\label{Dformula2}
\end{eqnarray}
\subsection{Example 1} 
In this subsection, we calculate the matrix element $\<>{I_3-J_3|D^1_{M_1M'_1}(\Omega)|I_3J_3}$ as an example.
Here and hereafter, we denote the nucleon state with the third component of the isospin $I_3$ and that of the spin $J_3$ simply by $|I_3J_3\rangle$.
This matrix element is given by the following integral,
\begin{eqnarray}
\<>{I_3-J_3|D^1_{M_1M'_1}(\Omega)|I_3J_3}
&\equiv&\frac{1}{(2\pi)^2}(-1)^{2I_3+1}\int d\Omega D^{1/2}_{-I_3-J_3}(\Omega)^*D^1_{M_1M'_1}(\Omega)D^{1/2}_{-I_3J_3}(\Omega).
\label{ex1}
\eey
Here, from \eq{Dformula1}, we note that
\begin{eqnarray}
D^1_{M_1M'_1}(\Omega)D^{1/2}_{-I_3J_3}(\Omega)
&=&\sum_{J=1/2}^{3/2}\<>{1{1\over2}M_1-I_3|JM}\<>{1{1\over2}M'_1J_3|JM'}D^{J}_{MM'}(\Omega)
\cr&=&\<>{1{1\over2}M_1-I_3|{1\over2}M}\<>{1{1\over2}M'_1J_3|{1\over2}M'}D^{1/2}_{MM'}(\Omega)
\cr&&+\<>{1{1\over2}M_1-I_3|{3\over2}M}\<>{1{1\over2}M'_1J_3|{3\over2}M'}D^{3/2}_{MM'}(\Omega).
\eey
We substitute this equation into \eq{ex1} and integrate the Euler angle using \eq{Dformula2}. The result reads as follows,
\begin{eqnarray}
\<>{I_3-J_3|D^1_{M_1M'_1}(\Omega)|I_3J_3}
&=&\<>{1{1\over2}M_1-I_3|{1\over2}M}\<>{1{1\over2}M'_1J_3|{1\over2}M'}
\delta(-I_3,M)\delta(-J_3,M')
\cr&=&(-1)^{1/2-I_3}{\sqrt{2}\over3}\delta(0,M_1)
[-\delta(-1,M'_1)\delta({1\over2},J_3)+\delta(1,M'_1)\delta(-{1\over2},J_3)].
\end{eqnarray}
\subsection{Example 2}
The second example is the matrix element of two $D$-functions: $\<>{I_3-J_3|D^1_{M_1M_1'}(\Omega)D^1_{M_2M_2'}(\Omega)|I_3J_3}$
given by
\bey
&&\<>{I_3-J_3|D^1_{M_1M_1'}(\Omega)D^1_{M_2M_2'}(\Omega)|I_3J_3}
\cr&=&\frac{1}{(2\pi)^2}(-1)^{2I_3+1}\int d\Omega D^{1/2}_{-I_3-J_3}(\Omega)^*D^1_{M_1M_1'}(\Omega)D^1_{M_2M_2'}(\Omega)D^{1/2}_{-I_3J_3}(\Omega).
\label{ex2}
\eey
Using \eq{Dformula1} two times, we obtain
\begin{eqnarray}
D^1_{M_1M_1'}(\Omega)D^1_{M_2M_2'}(\Omega)D^{1/2}_{-I_3J_3}(\Omega)
&=&\<>{11M_1M_2|00}\<>{11M'_1M'_2|00'}D^{1/2}_{-I_3J_3}(\Omega)
\cr&&+\<>{11M_1M_2|1M_3}\<>{11M'_1M'_2|1M'_3}
(\<>{1{1\over2}M_3-I_3|{1\over2}M}\<>{1{1\over2}M'_3J_3|{1\over2}M'}D^{1/2}_{MM'}(\Omega)
\cr&&+\<>{1{1\over2}M_3-I_3|{3\over2}M}\<>{1{1\over2}M'_3J_3|{3\over2}M'}D^{3/2}_{MM'}(\Omega))
\cr&&+\<>{11M_1M_2|2M_3}\<>{11M'_1M'_2|2M'_3}D^{2}_{M_3M'_3}(\Omega)D^{1/2}_{-I_3J_3}(\Omega).
\eey
We substitute this equation into \eq{ex2} and carry out the integration of the Euler angle using \eq{Dformula2} to obtain
\begin{eqnarray}
\<>{I_3-J_3|D^1_{M_1M_1'}(\Omega)D^1_{M_2M_2'}(\Omega)|I_3J_3}
&=&\<>{11M_1M_2|1M_3}\<>{11M'_1M'_2|1M'_3}
\cr&&\<>{1{1\over2}M_3-I_3|{1\over2}M}\<>{1{1\over2}M'_3J_3|{1\over2}M'}
\delta(-I_3,M)\delta(-J_3,M')
\cr&=&
(-1)^{1/2-I_3}{\sqrt{2}\over3}{1\over2}[\delta(M_1,1)\delta(M_2,-1)-\delta(M_1,-1)\delta(M_2,1)]
\cr&&[\{\delta(M'_1,1)\delta(M'_2,0)-\delta(M'_1,0)\delta(M'_2,1)\}\delta(-{1\over2},J_3)
\cr&&-\{-\delta(M'_1,-1)\delta(M'_2,0)+\delta(M'_1,0)\delta(M'_2,-1)\}\delta({1\over2},J_3)].
\end{eqnarray}

\def\Ref#1{[\ref{#1}]}
\def\Refs#1#2{[\ref{#1},\ref{#2}]}
\def\npb#1#2#3{{Nucl. Phys.\,}{\bf B{#1}},\,#2\,(#3)}
\def\npa#1#2#3{{Nucl. Phys.\,}{\bf A{#1}},\,#2\,(#3)}
\def\np#1#2#3{{Nucl. Phys.\,}{\bf{#1}},\,#2\,(#3)}
\def\plb#1#2#3{{Phys. Lett.\,}{\bf B{#1}},\,#2\,(#3)}
\def\prl#1#2#3{{Phys. Rev. Lett.\,}{\bf{#1}},\,#2\,(#3)}
\def\prd#1#2#3{{Phys. Rev.\,}{\bf D{#1}},\,#2\,(#3)}
\def\prc#1#2#3{{Phys. Rev.\,}{\bf C{#1}},\,#2\,(#3)}
\def\prb#1#2#3{{Phys. Rev.\,}{\bf B{#1}},\,#2\,(#3)}
\def\pr#1#2#3{{Phys. Rev.\,}{\bf{#1}},\,#2\,(#3)}
\def\ap#1#2#3{{Ann. Phys.\,}{\bf{#1}},\,#2\,(#3)}
\def\prep#1#2#3{{Phys. Reports\,}{\bf{#1}},\,#2\,(#3)}
\def\rmp#1#2#3{{Rev. Mod. Phys.\,}{\bf{#1}},\,#2\,(#3)}
\def\cmp#1#2#3{{Comm. Math. Phys.\,}{\bf{#1}},\,#2\,(#3)}
\def\ptp#1#2#3{{Prog. Theor. Phys.\,}{\bf{#1}},\,#2\,(#3)}
\def\ib#1#2#3{{\it ibid.\,}{\bf{#1}},\,#2\,(#3)}
\def\zsc#1#2#3{{Z. Phys. \,}{\bf C{#1}},\,#2\,(#3)}
\def\zsa#1#2#3{{Z. Phys. \,}{\bf A{#1}},\,#2\,(#3)}
\def\intj#1#2#3{{Int. J. Mod. Phys.\,}{\bf A{#1}},\,#2\,(#3)}
\def\sjnp#1#2#3{{Sov. J. Nucl. Phys.\,}{\bf #1},\,#2\,(#3)}
\def\pan#1#2#3{{Phys. Atom. Nucl.\,}{\bf #1},\,#2\,(#3)}
\def\app#1#2#3{{Acta. Phys. Pol.\,}{\bf #1},\,#2\,(#3)}
\def\jmp#1#2#3{{J. Math. Phys.\,}{\bf {#1}},\,#2\,(#3)}
\def\cp#1#2#3{{Coll. Phen.\,}{\bf {#1}},\,#2\,(#3)}
\def\epjc#1#2#3{{Eur. Phys. J.\,}{\bf C{#1}},\,#2\,(#3)}
\def\mpla#1#2#3{{Mod. Phys. Lett.\,}{\bf A{#1}},\,#2\,(#3)}
\def\etal{{\it et al.}}

\end{document}